\documentstyle[12pt]{article}
\renewcommand{\a}{\alpha}
\renewcommand{\b}{\beta}
\renewcommand{\d}{\delta}

\newcommand{\g}{\gamma}

\renewcommand{\l}{\lambda}
\newcommand{\m}{\mu}
\newcommand{\n}{\nu}

\renewcommand{\o}{\omega}

\newcommand{\r}{\rho}
\newcommand{\s}{\sigma}

\newcommand{\th}{\theta}
\newcommand{\vf}{{\varphi}}
\newcommand{\G}{\Gamma}

\newcommand{\ad}{{\dot{\alpha}}}
\newcommand{\dd}{{\,D\!D\,}}
\newcommand{\ddb}{{\,{\bar D}\!{\bar D}\,}}
\newcommand{\Db}{{\bar D}}

\newcommand{\thb}{{\bar\theta}}

\newcommand{\real}{{{\rm I} \kern -.19em {\rm R}}}
\newcommand{\tr}{{\rm {Tr} \,}}
\newcommand{\half}{\frac 1 2}
\newcommand{\pa}{\partial}

\newcommand{\intd}{{\int d^4 \! x \,}}

\newcommand{\disp}{\displaystyle}
\newcommand{\pad}[2]{{\disp{\frac{\partial #1}{\partial #2}}}}
\newcommand{\fud}[2]{{\disp{\frac{\delta #1}{\delta #2}}}}
\newcommand{\WWb}{{\bar{\cal W}}}
\newcommand{\complex}{{{\rm I} \kern -.59em {\rm C}}}
\newcommand{\naturel}{{{\rm I} \kern -.2em {\rm N}}}
\newcommand{\hc}{\ {\rm h.c.}\ }
\newcommand{\ddt}{{d^2 \theta \,}}

\newcommand{\ie}{{{\em i.e.},\ }}
\newcommand{\eg}{{{\em e.g.},\ }}

\newcommand{\NN}{{\cal N}}
\newcommand{\OO}{{\cal O}}

\newcommand{\SS}{{\cal S}}

\newcommand{\WW}{{\cal W}}

\newcommand{\Lp}{\displaystyle{\biggl(}}
\newcommand{\Rp}{\displaystyle{\biggr)}}

\newcommand{\lc}{\displaystyle{\bigl[}}
\newcommand{\rc}{\displaystyle{\bigr]}}
\newcommand{\Lc}{\displaystyle{\biggl[}}
\newcommand{\Rc}{\displaystyle{\biggr]}}

\newcommand{\la}{\displaystyle{\bigl\{}}
\newcommand{\ra}{\displaystyle{\bigr\}}}

\newcommand{\be}{\begin{equation}}
\newcommand{\ee}{\end{equation}}
\newcommand{\een}[1]{\label{#1}\end{equation}}
\newcommand{\beq}{\begin{eqnarray}}
\newcommand{\eeq}{\end{eqnarray}}
\newcommand{\eeqn}[1]{\label{#1}\end{eqnarray}}
\newcommand{\ba}{\begin{array}}
\newcommand{\ea}{\end{array}}
\newcommand{\brcl}{\begin{equation}\begin{array}{rcl}}
\newcommand{\ercl}{\end{array}\end{equation}}
\newcommand{\ercln}[1]{\end{array}\label{#1}\end{equation}}
\newcommand{\nn}{\nonumber}
\renewcommand{\=}{&=&} %seems not to work in footnotes
\newcommand{\qq}{&\qquad &}
\newcommand{\equ}[1]{(\ref{#1})}
\newcommand{\journal}[4]{{\sl #1~}{\bf #2},\,(19#3)\,#4;}
\newcommand{\hpa}{\journal {Helv. Phys. Acta}}

\newcommand{\np}{\journal {Nucl. Phys.}}

\makeatletter
\@addtoreset{equation}{section}
\makeatother
\renewcommand{\theequation}{\thesection.\arabic{equation}}
\catcode`\@=11
\def\marginnote#1{}
\newcount\hour
\newcount\minute
\newtoks\amorpm
\hour=\time\divide\hour by60
\minute=\time{\multiply\hour by60 \global\advance\minute by-\hour}
\edef\standardtime{{\ifnum\hour<12 \global\amorpm={am}%
    \else\global\amorpm={pm}\advance\hour by-12 \fi
    \ifnum\hour=0 \hour=12 \fi
    \number\hour:\ifnum\minute<10 0\fi\number\minute\the\amorpm}}
\edef\militarytime{\number\hour:\ifnum\minute<10 0\fi\number\minute}
\def\draftlabel#1{{\@bsphack\if@filesw {\let\thepage\relax
  \xdef\@gtempa{\write\@auxout{\string
   \newlabel{#1}{{\@currentlabel}{\thepage}}}}}\@gtempa
  \if@nobreak \ifvmode\nobreak\fi\fi\fi\@esphack}
    \gdef\@eqnlabel{#1}}
\def\@eqnlabel{}
\def\@vacuum{}
\def\draftmarginnote#1{\marginpar{\raggedright\scriptsize\tt#1}}
\def\draft{\oddsidemargin -.5truein
    \def\@oddfoot{\sl preliminary draft \hfil
    \rm\thepage\hfil\sl\today\quad\militarytime}
    \let\@evenfoot\@oddfoot \overfullrule 3pt
    \let\label=\draftlabel
    \let\marginnote=\draftmarginnote
  \def\@eqnnum{(\theequation)\rlap{\kern\marginparsep\tt\@eqnlabel}%
\global\let\@eqnlabel\@vacuum} }

\def\preprint{\twocolumn\sloppy\flushbottom\parindent 1em
    \leftmargini 2em\leftmarginv .5em\leftmarginvi .5em
    \oddsidemargin -.5in  \evensidemargin -.5in
    \columnsep 15mm \footheight 0pt
    \textwidth 250mmin   \topmargin -.4in
    \headheight 12pt \topskip .4in
    \textheight 175mm
    \footskip 0pt
    \def\@oddhead{\thepage\hfil\addtocounter{page}{1}\thepage}
    \let\@evenhead\@oddhead \def\@oddfoot{} \def\@evenfoot{} }

\def\titlepage{\@restonecolfalse\if@twocolumn\@restonecoltrue\onecolumn
   \else \newpage \fi \thispagestyle{empty}\c@page\z@
    \def\thefootnote{\fnsymbol{footnote}} }

\def\endtitlepage{\if@restonecol\twocolumn \else \fi
    \def\thefootnote{\arabic{footnote}}
    \setcounter{footnote}{0}} %\c@footnote\z@ }

\catcode`@=12
\relax
\begin{document}
%%%%%%%%%%%%%%%%%%%%%%%%%%%%%%%
%%%%%%%%%%%%%%%%%%%%%%%%%%%%%%%
%%%%%%%%%%%%%%%%%%%%%%%%%%%%%%%
%%%%%%%%%%%%%%%%%%%%%%%%%%%%%%%
\setlength{\baselineskip}{3.2ex}
%\setlength{\baselineskip}{5ex}
%%%%%%%%%%%%%%%%%%%%%%%%%%%%%%
\null April 1996\hfill NEIP-96-001\\
\null\hfill BUTP-96-7\\
\null\hfill hep-ph/9604216\\

\vspace*{-.1cm}

\begin{center}
{\Large\bf All-order Finiteness
in $N=1$ SYM Theories:}\\[2mm]
{\Large\bf Criteria and
Applications\footnote{\baselineskip=0pt  Invited paper,
Workshop on STU-Dualities and
Non-Perturbative Phenomena in
Superstrings and Supergravity,
CERN, Geneva, Nov 27 - Dec 1, 1995, to
appear in the proceedings.}$^,$\footnote{Work
supported in part by the Swiss
National Science Foundation, and
by C.E.U. projects SC1-CT91-0729; CHRX-CT93-0319.}}

\vspace*{4ex}

{Claudio LUCCHESI}

{\it Institut de Physique,
Universit\'e de Neuch\^atel}\\
{\it 1 rue Breguet, CH --
2000 Neuch\^atel (Switzerland)}\\

\vspace*{2ex}

{\small and}

{George ZOUPANOS\footnote{On leave
from Physics Department,
National Technical University, GR --
157 80 Zografou, Athens (Greece).}}

{\it Institute of Theoretical Physics,
University of Bern}\\
{\it Sidlerstrasse 5, CH -- 3012 Bern
(Switzerland)}\\

{\it and}\\
{\it Max-Planck Institut f\"ur Physik,
Werner Heisenberg Institut}\\
{\it F\"ohringer Ring 6,
D -- 80805 Munich (Germany)}\\

\vspace*{5ex}

{\bf Abstract}

\end{center}
\vspace*{-2ex}
{\small As a motivation, we first recall the possible connection of
electric-magnetic duality to finiteness in $N=1$ super-Yang-Mills theories
(SYM). Then, we present the criterion for all-order
finiteness (\ie vanishing of the  $\b$-functions at all orders) in $N=1$ SYM.
Finally, we apply this finiteness criterion to an $SU(5)$ SGUT. The latter
turns out to be all-order finite if one imposes additional symmetries.}
%%%%%%%%%%%%%%%%%%%%%%%
%%%%%%%%%%%%%%%%%%%%%%%
%%%%%%%%%%%%%%%%%%%%%%%
\vspace{-3ex}
\section{Introduction}
\label{sec0}
\vspace{-2ex}
Our aim is to present a criterion for all-order
 finiteness in $N=1$ SYM theories, to outline
 its derivation, and to exhibit an application
yielding an all-order finite supersymmetric GUT.
As a motivation, we would first like to
attempt at situating the discussion of all-order
finiteness within the larger and exciting
context of (electric-magnetic) duality in
supersymmetric gauge theories. Holomorphy
and duality in minimal and extended supersymmetry
\cite{holdual} form indeed an extremely active
domain of research,  to which the present
conference  is devoted.

We start, following \cite{bilal,olive}, by
recalling basic ideas about duality and
perturbative finiteness. The electric-magnetic
duality of Maxwell's equations in the vacuum
is a symmetry under the exchange of the
electric and magnetic fields. In the presence
of sources, one is led to introduce magnetic
monopoles with magnetic charges $q_m$, in
addition to the electric charges $q_e$, which obey
 Dirac's charge quantization condition
$q_e\, q_m =2\pi\,n,\ n\in\naturel$. For the
elementary charges ($n=1$), duality therefore
exchanges $q_e$ with its inverse
(up to the $2\pi$ factor) $2\pi/q_e=q_m$.

The electric-magnetic duality of Maxwell's theory
motivates one to search for an analogous
symmetry in the realm of quantum field theory,
\eg QED. The latter, supporting no magnetic
monopoles, has to be rejected, and one
therefore turns to spontaneously broken
Yang-Mills theories. In that framework,
due to the running of
the couplings, duality can only be established
if it can be made to hold at any scale $\m$,
that is, the symmetry under the exchange of $\l(\m)$
with the dual coupling $\sim 1/\l(\m)$ has to
be scale-independent. The latter can be achieved
 provided the couplings  {\it do not run}, \ie provided
their $\b$-functions vanish exactly. This is the
 case in $N=4$ supersymmetric  Yang-Mills theories,
for which Montonen and Olive
\cite{montol}  have conjectured
 that electric-magnetic duality might be an exact
symmetry. Similarly, $N=2$ SYM theories can be
made finite  by choosing appropriate combinations
of gauge group and matter fields which lead to
vanishing one-loop $\b$-functions. A non-renormalization
 theorem then guarantees that the $\b$-functions
vanish  above one-loop \cite{hsw}. The discussion
of perturbative finiteness in $N=2,4$ SYM is hence seen to
be closely related to that of electric-magnetic duality.

There is an obvious interest in discussing
the case of $N=1$ SYM theories as well.
In contrast to $N=4$ or $N=2$, the $N=1$ case
is relevant to the supersymmetric GUT's
of low-energy phenomenology,
as those yielding the minimal supersymmetric standard
model (MSSM).
Therefore, in the same spirit of relating perturbative
finiteness to electric-magnetic duality, we focus
in the present paper on the
case of minimal, $N=1$ supersymmetry. We present the criterion
for all-order finiteness in $N=1$ SYM of \cite{pisi,LPS}, give
 an outline of the proof and produce an example of application.

The paper is structured as follows.
In Section \ref{sec1}, after reviewing the criteria
for one- and two-loop finiteness in $N=1$ SYM
theories \cite{PW}, we show how lower-orders
finiteness can be extended to all orders
\cite{pisi,LPS}, by imposing, as a consistency
condition for higher orders, that the
gauge and Yukawa couplings obey reduction in
the sense of Oehme and Zimmermann \cite{Zim}.
All-order  scale invariant theories  hence possess
a single independent coupling,
which does not run.  It should be noted, in the
context of electric-magnetic duality, that the proof
of this finiteness criterion for $N=1$ SYM is a
rigorous extension of a formal argument proposed
for the finiteness of $N=4$ \cite{argn=4}.

In Section \ref{sec2}, we present the {\it criterion
 for all-order vanishing  $\b$-functions in $N=1$
SYM} of \cite{LPS}, an exact result which is based
on hypotheses operating exclusively at
the one-loop level. We start by recalling the structure
of the supercurrent multiplet anomaly
\cite{ferrarazumino,piguetsiboldsc,PSbook}, which
yields an important relation \cite{pisi} connecting the
conformal anomalies, \ie the $\b$-functions and the
anomalous dimensions, to the  axial and $R$-axial
anomalies. The latter  axial and $R$-axial anomalies
being non-renormalized,  they are given by their
one-loop values. Vanishing  of the latters is among
the hypotheses of the finiteness criterion. The consistency
requirement of couplings reduction translates  into a
further hypothesis on the unicity of the solution to the
conditions of vanishing one-loop Yukawa $\b$-functions.

It turns out that fulfilling the criterion generally
means imposing global, Lie-type or discrete symmetries
that restrict the superpotential. As an illustration
we present, in Section \ref{sec3},
among the possible all-order finite $N=1$ SYM models,
a realistic supersymmetric $SU(5)$ gauge theory with
discrete symmetries \cite{kmz}. The latter finite model
can be tested phenomenologically, based on its
prediction for the top quark mass, which is in
agreement with experimental data.

To conclude this Introdution, let us mention
that there exists related approaches to all-order
finiteness in $N=1$ SYM, as those in refs.
\cite{ermushev,DJZ,strassler}.
The work of Intrilligator, Leigh and Strassler \cite{strassler}
is of special relevance here due to
its implications for duality symmetry.

%%%%%%%%%%%%%%%%%%%%%
%%%%%%%%%%%%%%%%%%%%%
%%%%%%%%%%%%%%%%%%%%%
\vspace{-3ex}
\section{Finite $N=1$ Supersymmetric Gauge Theories}
\vspace{-2ex}
\label{sec1}
We consider a chiral, anomaly free,
$N=1$ globally supersymmetric
Yang-Mills theory based on a group G with gauge coupling
 $g$, and with the superpotential
\beq
W(\phi^i)&=& \frac{1}{2}\,m_{ij} \,\phi^{i}\,\phi^{j}+
\frac{1}{6}\,\l_{ijk}\,\phi^{i}\,\phi^{j}\,\phi^{k}~,
\label{supot}
\eeq
where $m_{ij}$ and $\l_{ijk}$ are gauge invariant tensors and
the matter fields $\phi^{i}$ transform
according to the irreducible representation $R_{i}$
of $G$. The
renormalization constants associated with the
superpotential \equ{supot}, assuming that
supersymmetry is preserved, are
\be
\phi^{0\,i}=(Z^{i}_{j})^\half\,\phi^{j}~,\qquad
m_{ij}^{0}=Z^{i'j'}_{ij}\,m_{i'j'}~,\qquad
\l_{ijk}^{0}=Z^{i'j'k'}_{ijk}\,\l_{i'j'k'}~.
\ee
The non-renormalization theorem for $N=1$ SYM \cite{nonre}
ensures the absence of mass and cubic infinities, therefore
\be
Z_{ijk}^{i'j'k'}\,(Z^{i''}_{i'})^\half\,(Z^{j''}_{j'})^\half
\,(Z^{k''}_{k'})^\half=\delta_{(i}^{i''}
\,\delta_{j}^{j''}\delta_{k)}^{k''}~,\qquad
Z_{ij}^{i'j'}\,(Z^{i''}_{i'})^\half\,(Z^{j''}_{j'})^\half
=\delta_{(i}^{i''}\,\delta_{j)}^{j''}~.
\ee
As a result the only surviving possible infinities are
those associated to the wave-function renormalization constants
$Z^{j}_{i}$, i.e., one infinity
for each field. The one-loop gauge
$\beta$-function is given by \cite{PW,barb}
\be
\beta^{(1)}_{g}~=~\frac{d g}{d\, {\rm ln}\mu} ~=~
\frac{g^3}{16\pi^2}\,\left[\,\sum_i\,T(R_{i})-3\,C_{2}(G)\,\right]~,
\label{betag}
\ee
where $\mu$ is the renormalization scale,
$T(R_{i})$ is the Dynkin index of $R_{i}$ and $C_{2}(G)$
is the quadratic Casimir of $G$. The one-loop $\beta$-functions
$\beta_{ijk}^{(1)}$ of the Yukawa couplings $\l_{ijk}$ are related
to the matrix $\gamma^{i\ (1)}_{\ j}$ of one-loop anomalous
dimensions of the matter fields $\phi^{i}$ as
\be
\beta^{(1)}_{ijk} ~=~
 \frac{d \l_{ijk}}{d\, {\rm ln}\,\mu}~=~\l_{ijl}\,\gamma^{l\ (1)}_{\ k}+
 \l_{ikl}\,\gamma^{l\ (1)}_{\ j}+
 \l_{jkl}\,\gamma^{l\ (1)}_{\ i}~,
\label{betay}
\ee
with \cite{PW,barb}
\be
\gamma^{i\ (1)}_{\ j}=\frac{1}{32\pi^2}\,[\,
\l^{ikl}\,\l_{jkl}-2\,g^2\,C_{2}(R_{i})\delta^i_{j}\,]\ ,
\label{gamay}
\ee
where $C_{2}(R_{i})$ is the quadratic Casimir of the irrep.
$R_{i}$, and $\l^{ijk}=\l_{ijk}^{*}$.

Necessary and sufficient conditions for one-loop finiteness
result from demanding that the one-loop gauge $\beta$-function
 \equ{betag}, respectively the one-loop
matter fields anomalous dimensions \equ{gamay}, vanish, \ie
\beq
\sum_i T(R_i) \= 3\, C_2(G)\ ,\label{1st}\\[2mm]
\l^{ikl} \l_{jkl} \= 2\,\, g^2 \, C_2(R_i)\,\delta ^i_j \ .
\label{2nd}
\eeq
These one-loop
finiteness conditions are known to be necessary
and sufficient for finiteness at the two-loop level \cite{PW}.
In case supersymmetry is broken by soft terms,
one-loop finiteness of the soft sector imposes
further constraints on it \cite{soft}.
In addition, the same set of
conditions that are sufficient for one-loop
finiteness of the soft breaking terms renders
the soft sector of the theory two-loop
finite \cite{jones}.

The one- and two-loop finiteness conditions
\equ{1st}-\equ{2nd} restrict considerably
the possible choices of the
irreps. $R_i$ for a given group $G$ as
well as the Yukawa couplings
in the superpotential (\ref{supot}).
Note in particular that the
finiteness conditions cannot be
applied to the supersymmetric standard model (SSM), since
the presence of a $U(1)$ gauge group is
incompatible with the condition (\ref{1st}), due to $C_2[U(1)]=0$.
This naturaly leads to the expectation that
finiteness should be attained at the
grand unified level only, the SSM
being just the corresponding
low-energy, effective theory.

Another important consequence of
one- and two-loop finiteness
is that supersymmetry (most probably) can only
be broken by soft breaking terms. Indeed,
due to the unacceptability of gauge singlets,
F-type \cite{f} spontaneous supersymmetry
breaking terms are incompatible with
finiteness, as well as D-type \cite{d}
spontaneous breaking which requires
the existence of a $U(1)$ gauge group.

A natural question to ask is what happens at higher loop orders.
The answer is contained in a theorem \cite{LPS}
which states the necessary and
sufficient conditions to achieve finiteness at all orders.
Before we formulate the
theorem let us make some introductory
remarks. The one and two-loop
finiteness conditions \equ{1st}-\equ{2nd} restrict the possible
choices of matter representations and impose relations between
 the gauge and Yukawa couplings
 [see \equ{2nd}]. To require such relations which
render the couplings mutually dependent, at a given
renormalization point, is trivial.
What is not trivial is to guarantee that relations leading to a
reduction of the couplings hold at any renormalization
point\footnote{ A recent paper \cite{JJN} deals with the possibility
of extending at all orders the fixed points of $\l_{ijk}/g$ (a simple
form of reduction of the couplings), by discussing lower-order
results.}. The necessary, but also sufficient,
condition for this to happen is to
require that such relations are solutions to
the reduction equations of Oehme and Zimmermann \cite{Zim},
\be
\b_g \,{{d \l_{ijk}\over dg}}=\b_{ijk}\ ,
\label{redred}
\ee
and hold at all orders.
It is remarkable that the existence of all-order solutions to
\equ{redred} can be decided at the one-loop level \cite{Zim}.
We shall come back to this point in Section \ref{sec2}.

Attempts at reducing
the couplings can be found in refs.
\cite{ccred1,ccred2,ccred3,otherguts}.
These models, in particular the ones in ref. \cite{otherguts},
as well as the finite model that
will be described in Section \ref{sec3},
predict among other things a window for
the top quark mass \cite{top,otherguts}.
More accurate measurements of the latter
will decide on their validity.

We now return to the all-order finiteness theorem of \cite{LPS}.
It is based on {(a)} the structure of the supercurrent in $N=1$ SYM
\cite{ferrarazumino,piguetsiboldsc,PSbook}, on {(b)} the
non-renormalization properties of $N=1$ chiral anomalies
\cite{pisi,LPS}, and finally on {(c)} the requirement of reduction
of the couplings, in the sense discussed above. The theorem
states that for an $N=1$ supersymmetric gauge theory based
on a simple gauge group (with representations
that are free of gauge anomalies),
the necessary and sufficient conditions for
$\beta_g$ and $\beta_{ijk}$ to vanish at all orders are the following:
{(i)} the one-loop finiteness conditions \equ{1st}-(\ref{2nd}) hold;
{(ii)} the reduction equations \equ{redred} admit a formal power
series solution; {(iii)} the latter,
in its lowest order, is also a solution
of the condition (\ref{2nd}).
Since, as mentioned above, the existence of all-order
solutions to the reduction equations can be decided
at the one-loop level, the theorem can be recast
in the form of a {\it criterion for all-order finiteness} \cite{LPS},
which is {\it based on exclusively one-loop hypotheses}. We
shall develop that formulation in the next section.

%%%%%%%%%%%%%%%%%%%%%%%%%
%%%%%%%%%%%%%%%%%%%%%%%%%
%%%%%%%%%%%%%%%%%%%%%%%%%
\vspace{-3ex}
\section{Criterion for All-order Vanishing $\b$-Functions}
\label{sec2}
\vspace{-2ex}
We start by describing formally, within the
``algebraic renormalization" approach \cite{pisor},
\ie through renormalized Ward identities, the
action for $N=1$ SYM (with a simple gauge group $G$)
and its symmetries. In order to avoid problems with
regularization, the theory is assumed to be renormalized
according to the superspace renormalization scheme of
\cite{PSbook}.

The real gauge and chiral matter
superfields are resp. denoted by $V$ and $\phi^i$.
The gauge-invariant superfield action reads (in a notation that
differs from the one used in the original literature, references
\cite{LPS, pisi}):
\beq
S^{\ \rm invariant}\=-{1\over 128 g^2}\tr
\left(\intd \ddt W^\a W_\a + \hc \right)+
{1\over 16}\ \intd d^4 \theta \,
\phi_i^\dagger \,(e^{V})^i_{\, j}\, \phi^j\nn\\[2mm]
&&+{1\over 6}\left(\intd \ddt \l_{ijk}
\, \phi^i\phi^j\phi^k + \hc\right)\ ,
\label{invaction}
\eeq
where we have omitted the supersymmetric mass terms of the
superpotential \equ{supot}. In other words, we consider all fields to
 be massive, but we treat the theory only asymptotically at large
Euclidean momenta. We therefore avoid all infrared problems
associated with the dimension zero of the vector superfield $V$.
[The construction of $N=1$ SYM theories with
{\it supersymmetry breaking} masses, using
the algebraic renormalization
approach (in the Wess-Zumino gauge)
developed in \cite{algrenorm},
has been addressed recently \cite{nos}$\,$].

The detailed gauge-fixing of
the action \equ{invaction} \cite{PSbook}
is beyond our purpose. Following
the BRS quantization procedure, one usually constructs the
generating functional of one-particle
irreducible Green's functions
(the classical action) as
\be
S_{\rm cl} = S^{\ \rm invariant} + S^{\ \rm gauge\ fixing} +
S^{\ \rm Faddeev-Popov}+ S^{\ \rm external\ sources}\ ,
\ee
and defines the quantum theory generating functional
\be
\G=S_{\rm cl}+\OO(\hbar^n)
\ee
to be the most general solution of a set of renormalized constraints
given by the gauge condition, the equations of motion, the rigid
and BRS symmetries, etc. A subset of these constraints is relevant
to the present context:

\noindent
1. $R$-symmetry. On a generic superfield
 $\vf=\phi^i\ ,\phi_i^\dagger\ ,V$, ghosts, antighosts,
{\it etc.}, $R$-trans\-for\-ma\-tions act infinitesimally as
\be
\d_R\vf=i\,( n_\vf+\th^\a\pa_{\th^\a}-\thb^\ad\pa_{\thb^\ad})\,\vf\ ,
\ee
with $R$-weights $n_\vf$ given by
$n_{\phi^i}=-n_{\phi_i^\dagger}=-{{2\over 3}}$, and all
other $n_{\vf}={0}$. The functional $R$-Ward identity reads
\be
\WW_R\G \equiv -i\sum_\vf\int \d_R\vf\ \fud{\G}{\vf}\simeq 0\ ,
\label{rinvward}
\ee
where the symbol $\simeq$ denotes equality up to soft breakings
 (\ie breaking terms that vanish in the deep Euclidean region)
of $R$-symmetry induced by supersymmetric masses, and
the convenient superspace integration
 measure is subsumed in $\int$.

\noindent
2. Supersymmetry, expressed through the usual (unbroken)
Ward identities
\be
\WW_\a \G = 0\ ,\qquad\WWb_\ad \G = 0\ .
\label{susyward}
\ee

\noindent
3. BRS invariance, acting as, \eg
\be
s\,e^V =e^V c_+ -{\bar c}_+ e^V\ ,\qquad
s\,\phi^i=-c_{+\,a}(T^a_R)^i_{\ j}\phi^j\ ,\qquad
s\,c_+=-\half\la c_+;c_+\ra\ ,
\label{ssss}
\ee
where $c_+,\ {\bar c}_+$ denote Faddeev-Popov ghosts and antighosts.
BRS invariance is encoded in a (non-linear) Ward identity, the
Slavnov identity \be
\SS(\G) \equiv -i \int \left[
\sum_{\ \vf\,|\,s\vf\ {\rm non-linear}}
\fud{\G}{Y_\vf}\fud{\G}{\vf}+
\sum_{\vf\,|\,s\vf\ {\rm linear}}
s\,\vf\,\fud{\G}{\vf}
\right]
\simeq 0\ ,
\label{slavnov}
\ee
where the $Y_\vf$'s are external
sources coupled to the non-linear
BRS variations $s\,\vf$, \eg those displayed in \equ{ssss}.
The corresponding
$\int s\vf\,Y_\vf$ terms in $S^{\ \rm external\ sources}$ make
it possible to define in the quantum theory the composite BRS
variations $s\vf$ as functional derivatives
of $\G$ w.r.t. the
sources $Y_\vf\ $. The Slavnov identity
\equ{slavnov} is satisfied provided
there is no gauge anomaly \cite{PSbook,chiralanomaly}. The
symbol $\simeq$ means that \equ{slavnov}, similarly to
 \equ{rinvward}, holds up to soft breakings of BRS invariance
induced by supersymmetric masses.

\noindent
4. A possible set of rigid chiral symmetries acting solely on the
matter superfields,
\be
\d_a\phi^i=i\,e_{a\ j}^{\ i}\phi^j\ ,\qquad
\d_a\phi_i^\dagger=-i\phi_j^\dagger\,e_{a\ i}^{\ j}\ ,\qquad
\d_a V=\d_ac_+=\d_a{\bar c}_+=\dots =0\ ,
\ee
generated by Hermitean charges $e_a=e_a^\dagger$.
The ``chiral" Ward identity
\be
\WW_a\G \equiv -i\int \sum_\vf \d_a\vf\,\fud{\G}{\vf}
=\sum_{i,j} e_{a\ j}^{\ i}
\left[\intd\ddt \phi^i\fud{}{\phi^j} - \hc\right]\G
\simeq 0
\label{chirsim}
\ee
is satisfied, up to soft breakings of
 chiral symmetry due to supersymmetric
masses [see also \equ{rinvward} and \equ{slavnov}], provided
$\l_{ijl}\, e_{a\ k}^{\ \,l} + \l_{jkl}
\, e_{a\ i}^{\ \,l} +\l_{kil}\, e_{a\ j}^{\ \,l}=0$.

%%%%%%%%%%%%%%%%%%%%%%%%%%%
The Ward operators for supersymmetry, translations and
$R$-transformations close under supersymmetry. As a consequence,
a superfield Ward operator $\hat\WW$ can be constructed out of the
Ward operators for $R$-transformations $\WW_R$, supersymmetry
$\WW_\a,\ \WWb_\ad$, and translations $\WW_\m^T$, as
\be
\hat\WW=\WW_R-i\,\th^\a\WW_\a + i\,\thb^\ad\WWb_\ad
-2\,\th^\a\s^\m_{\a\ad}\thb^\ad\,\WW_\m^T \ ;
\label{ww}
\ee
the Ward identities \equ{rinvward}, \equ{susyward} are thus comprised
in
\be
\hat\WW\G\simeq 0\ .
\ee
The component N{\oe}ther currents associated to the symmetries
in \equ{ww}, \ie the $R$-current $R_\m$, the supersymmetry
currents $Q_{\m\a},\ {\bar Q}_{\m\ad}$, and the energy-momentum
tensor $T_{\m\n}$, form a supermultiplet -- the supercurrent
\cite{ferrarazumino,piguetsiboldsc} $V_\m$
\be
V_\m(x,\th,\thb)=R_\m(x)-i\,\th^\a Q_{\m\a}(x) + i\,\thb^\ad
{\bar Q}_{\m\ad}(x) - 2(\th^\a\s^\n_{\a\ad}\thb^\ad)\,T_{\m\n}(x)
+ \dots\ ,
\ee
with $V_{\a\ad}=\half \s^\m_{\a\ad}\,V_\m$ obeying the supertrace
identity \cite{piguetsiboldsc}, which we write schematically as
\be
\Db^\ad V_{\a\ad}\sim 2\,D_\a{\bf S}+{\rm c.t.}\ .
\label{sutrace}
\ee
``c.t." denotes the
relevant contact terms. Applying
$D^\a=\pa_{\theta_\a}-i\s_\m^{\a\ad}
\thb_\ad\,\pa^\m$ to \equ{sutrace}
and $\Db^\ad=-\pa_{\bar\theta_\ad}
+i\theta_\a\s_\m^{\a\ad}\,\pa^\m$ to
the conjugate equation, and combining,
one gets the conservation
condition for the supercurrent
\be
\pa^\m V_\m\sim i\,(\dd{\bf S}-\ddb{\bf\bar S})+{\rm c.t.}\ .
\label{Vcons}
\ee
In \equ{sutrace} and \equ{Vcons}, ${\bf S}$ is a chiral superfield
insertion of dimension 3 and $R$-weight $-2$ known as the
supercurrent anomaly \cite{piguetsiboldsc}. It has the structure
of a supermultiplet containing the anomalies of the supercurrent
components, \ie the anomalous divergence of the $R$-current,
the ($\g$-)traces of the supersymmetry currents, and the trace of
the energy-momentum tensor.
Indeed, in the lowest component, \equ{Vcons} yields an Abelian
chiral anomaly breaking the $R$-current divergence
\be
\pa_\m R^\m \sim i\,(\dd{\bf S}-\ddb{\bf\bar S})+{\rm c.t.}\ .
\label{anomrdiv}
\ee
It is remarkable that the supertrace identity \equ{sutrace}
also yields dilatation anomalies in
 the energy-momentum trace
\be
T^\m_{\ \m} \sim \o_D\G
-{3\over 2}\,(\dd{\bf S}+\ddb{\bf\bar S})\ ,
\label{anomemtrace}
\ee
where  $\o_D$ is just, upon integration, the Ward operator of dilatations
\be
\int \o_D \equiv \WW_D \equiv -i\sum_\vf\int \d_D\vf\ \fud{}{\vf}\ ,
\label{warddil}
\ee
with the infinitesimal transformation law
\be
\d_D\vf=( d_\vf+x^\m\pa_\m
+\half\th^\a\pa_{\th^\a}
+\half\thb^\ad\pa_{\thb^\ad})\ \vf\ .
\ee

Our task is now to relate the Abelian chiral anomaly in the
$R$-current divergence \equ{anomrdiv} and the dilatation
anomalies in the energy-momentum trace \equ{anomemtrace}
to the Abelian anomalies associated to the (possible) chiral
symmetries $\WW_a$ \equ{chirsim}. The natural setting for
deriving such a relation is provided by the
Callan-Symanzik equation. We shall arrive at its formulation
 by expanding the supercurrent anomaly $\bf S$ in a basis of
dimension 3, BRS-invariant, chiral insertions $\{L_n\}$ as
\be
{\bf S} =\b_g\,L_g +\sum_{\l_{ijk}} \b_{ijk}\,L_{ijk}
 -\sum_{i,j}\g^j_{\ i}\,L^i_{\ j} + \dots\ ,
\label{sexp}
\ee
where the dots stand for insertions which are not essential
in the present context. The $L_n$'s (the choice of which
shall be justified by the Callan-Symanzik equation)
are defined in terms of derivatives of the action functional
w.r.t. the gauge and Yukawa couplings,
\be
\pa_g\G \equiv \intd \ddt L_g\ + \hc \ ,\quad
\pa_{\l_{ijk}}\G \equiv \intd \ddt L_{ijk}\ + \hc \ ,
\label{ln1}
\ee
and in terms of the ``counting operators" $\NN^i_{\ j}$, defined as
\be
\NN^i_{\ j}\G \equiv \left[\intd \ddt \phi^i{\d\over\d \phi^j}\
+\hc\right]\G\equiv \intd \ddt L^i_{\ j}\ + \hc
\ .
\label{SL}
\ee
Let us replace into the energy-momentum trace
\equ{anomemtrace} the expansion for $\bf S$
\equ{sexp} and the definitions of the $L_n$'s \equ{ln1},
\equ{SL}. Then, relating the (broken) Ward identity of
dilatations to the scaling operator through the dimensional
analysis identity $\WW_D\G=\m\,\pa_{\m}$, one arrives at the
Callan-Symanzik equation
\be
\Lc \ \m\pa_{\m} +\b_g\,\pa_g
+\sum_{\l_{ijk}} \b_{ijk}\,\pa_{\l_{ijk}}
-\sum_{i,j}\g^j_{\ i}\,\NN^i_{\ j}
+ \dots \Rc \, \G =0\ .
\label{CS}
\ee
The latter describes how dilation invariance is broken by
the $\b$-functions $\b_g$, $\b_{ijk}$ associated to the
renormalization of the gauge, resp. Yukawa couplings,
and by the anomalous dimensions $\g^i_{\ j}$.
Eq. \equ{CS} justifies
{\it a posteriori} the choices of the coefficients in \equ{sexp},
and of the definitions \equ{ln1}, \equ{SL}.

We now perform a change of basis for the counting
operators $\NN^i_{\ j}$ \equ{SL},
\be
\la \NN^i_{\ j} \ra \longrightarrow
\la \NN_{0a}\equiv e_{0a\ i}^{\ \ j}\
\NN^i_{\ j}\ra \oplus \la \NN_{1k}\ra\ ,
\label{chbasis}
\ee
where the $e_{0a\ j}^{\ \ i}$ are charge matrices
corresponding to the center of the algebra
$\la\WW_a\ra$ formed by the chiral symmetries
 \equ{chirsim}, \ie
\be
\WW_{0a}\G \equiv\sum_{i,j} e_{0a\ j}^{\phantom{0}\ i}
\left[\intd\ddt \phi^i\fud{}{\phi^j} - \hc\right]\G
\simeq 0
\label{chirsima}
\ee
with
\be
\lc \,\WW_{0a}\,;\WW_b\,\rc \ =\ 0, \qquad\forall b\ .
\ee
The new counting operators
$\NN_{0a}\equiv e_{0a\ i}^{\ \ j}\ \NN^i_{\ j}$ \equ{chbasis}
annihilate the superpotential in the (asymptotic)
action \equ{invaction}, in the sense that
\be
\NN_{0a} \left[\intd\ddt \l_{ijk}\,\phi^{i}\,\phi^{j}\,\phi^{k}\
+\hc\right] =0\ .
\ee

One can show that the supercurrent anomaly $\bf S$, as
well as each of the insertions of its expansion in the new
basis $\{\NN_{0a},\NN_{1k}\}$ \equ{chbasis}, omitting the
unessential terms $\NN_{1k}$,
\be
{\bf S} =\b_g\,L_g +\sum_{\l_{ijk}} \b_{ijk}\,L_{ijk}
-\sum_{a}\g_{0a}\,L_{0a} + \dots\ ,
\label{sexpa}
\ee
can be expressed as the $K_3^0$-dependent terms,
\be
{\bf S} =\ddb (r \, K_3^0 + \dots)\ ,
\label{325}
\ee
and
\be
L_g =\ddb \left({1\over 128 g^3} + r_g \right)
\, K_3^0 + \dots\ ,\quad
L_{ijk} =\ddb r_{ijk} \, K_3^0 + \dots\ ,\quad
L_{0a} =\ddb r_{0a} \, K_3^0 + \dots\ ,
\label{326}
\ee
where $K_3^0$ is the ghost number zero, supersymmetric
Chern-Simons term, and the dots stand for unessential terms.
Replacing these expressions into \equ{sexpa}, and identifying
the coefficients of the $K_3^0$-dependent terms, yields an
 important relation \cite{pisi}
among the coefficients of the anomaly expansion \equ{sexpa}
 and those of \equ{325}, \equ{326}:
\be
r=\b_g\ \Lp {1\over 128\, g^3} + r_g\Rp
+\sum_{\l_{ijk}}\b_{ijk}\,r_{ijk}
-\sum_a \g_{0a}\, r_{0a}\ ,\label{rrr}
\ee
where $r_g$, $r_{ijk}$ are of order $\hbar$ at least, whereas $r$ and $r_{0a}$
are
strictly of order $\hbar$.

Indeed,
specializing to the case under consideration,
the non-renormalization
 theorem for chiral anomalies in $N=1$ SYM
(see \cite{pisi,LPS})
tells us that $r$ and $r_{0a}$ in \equ{rrr}
are non-renormalized, \ie
they are of order one in $\hbar$.
$r$ is the coefficient of the Abelian
anomaly of the $R$-axial current,
and the $r_{0a}$'s are the coefficients
of the Abelian anomalies of
the axial currents associated to the
chiral $\WW_{0a}$-symmetries
 \equ{chirsima}. $r$ and $r_{0a}$ are
proportional to their one-loop
values \cite{PW,pisi,LPS}
\be
r \sim \b^{(1)}_g
\sim\sum_i T(R_i)-3\,C_2(G)\ ,\qquad
r_{0a} \sim \sum_i e_{a\ i}^{\ j} \,T(R_i)\ .
\label{rr0a}
\ee
Note that at the order one in $\hbar$, \equ{rrr} reduces
to $ \b^{(1)}_g=128\,g^3 r$, which just
corresponds to the first ``$\sim$" in eqs.
\equ{rr0a}, up to the coefficient.

The proof of the non-renormalization theorem \cite{pisi,LPS}
uses the fact that the supersymmetric Chern-Simons three-form
 $K^0_3$ is related through the supersymmetric descent equations
 to the zero-form $K^3_0 = {1 \over 3}\ \tr\, c_+^3$,
the cubed ghost field
insertion. The non-renormalization theorem for chiral vertices
\cite{fula,PSbook} guarantees the finiteness of the latter insertion.
(Another derivation of the finiteness of $\tr\, c_+^3$, which
is based on the supersymmetric antighost equation, has been given
in \cite{olsi}).

%%%%%%%%%%%%%%%%%%%%%%%%%%
We now state the criterion for
all-order vanishing $\b$-functions,
in the form announced at the end of Section \ref{sec1}.

\noindent{\bf Criterion for all-order
vanishing $\b$-functions:}\\[2mm]
Consider an N=1 super-Yang-Mills
theory with simple gauge group. {\bf If}\\
{\bf (i)} there is no gauge anomaly,\\
{\bf (ii)} the gauge $\b$-function vanishes at one loop [eq. \equ{1st}],
\be
\b_g^{(1)}=0\ ,
\ee
{\bf (iii)} there exist solutions of the
form $\l_{ijk} =\r_{ijk}\,g,\ \r_{ijk}\in\complex$, to the
conditions of vanishing one-loop matter fields
anomalous dimensions [eqs. \equ{2nd}]
\be
\g^{i\ (1)}_{\ j} =0\ ,
\label{410}
\ee
and {\bf (iv)} these solutions are isolated and non-degenerate when
considered as solutions of the conditions of vanishing one-loop
Yukawa $\b$-functions [see eq. \equ{betay}]
\be
\b^{(1)}_{ijk} =0\ ,
\label{411}
\ee
{\it {\bf then} each of the solutions $\l_{ijk} =\r_{ijk}\,g$ can be uniquely
extended to formal power series in $g$, and the associated SYM models
depend on a single coupling constant
(\eg the gauge coupling $g$)
with a $\b$-function which vanishes at all orders}.

Some comments are in order. By ``isolated", we
mean that the solutions cannot be multiple zeroes,
whereas by ``non-degenerate" we forbid parametric
 families of solutions. Indeed,
the solutions of  $\g^{i\ (1)}_{\ j} =0$ are generally
multiple zeroes or come in one-parameter families.
To obtain a SYM model with one isolated and
non-degenerate solution (\ie a unique solution
for that model), one generally needs to restrict
the superpotential by imposing global, chiral or
discrete, symmetries. One solution of $\g^{i\ (1)}_{\ j} =0$
which is unique when regarded as a solution of
$\b^{(1)}_{ijk} =0$ therefore corresponds, if it exists,
 to a given global symmetry of the superpotential.
This is the meaning
of the finiteness criterion: there can be different,
arbitrarily multiple or degenerate solutions to
$\g^{i\ (1)}_{\ j} =0$. Each of them {\it may} yield
a finite SYM model with global symmetries, assuming
that such symmetries exist. If more than one finite
model can be constructed for a given unconstrained
$N=1$ SYM theory, then each of these models
corresponds to the original theory with an additional
global symmetry specific to that model.

Note that, due to the fact that the couplings are complex,
there generally are undetermined phases left in the
solution of  $\b^{(1)}_{ijk} =0$, hence the latter is a
parametric family. These phases $\varphi$  can be
set to zero by hand provided the corresponding
$\b$-functions, schematically
$\b_\varphi={\rm Im}\,(\b_\l / \l)$, vanish.
$\b_\varphi =0$ holds if one uses a renormalization
scheme that preserves at all orders the one-loop relation
\equ{betay} between the $\b$- and $\g$-functions. For
details, see \cite{LPS}.

The conditions $\b_g^{(1)}=\g^{i\ (1)}_{\ j} =0$ have been  known
for some time to guarantee one- and
two-loop vanishing of the $\b$-functions
\cite{PW} (see Section \ref{sec1}). Models which
fulfill these conditions are tabulated
in \cite{HPS}, for the most popular (simple) gauge groups.
Conditions (iii) and (iv)
represent therefore consistency requirements
that are necessary in order to extend the vanishing of the
$\b$-functions at all orders.

Some models satisfying the all-order finiteness criterion
are known. An all-order finite $SU(6)$ SYM theory has
been presented in \cite{LPS}. Other attempts at finding
all-order finite models have resulted in constraining the
initial theory by imposing discrete (orbifold-type) symmetries
\cite{kmz}; an example is presented in
Section \ref{sec3} below.

Let us now sketch the proof of the finiteness criterion.
With the expressions for $r$ and $r_{0a}$
\equ{rr0a}, it follows from (ii) and (iii) that
\be
r=0\ , \quad{\rm resp.}\ r_{0a}=0\ ,
\label{000}
\ee
the latter resulting from a corollary to the non-renormalization
theorem for $N=1$ SYM stated above: the conditions
$\g^{i\ (1)}_{\ j}=0$ are compatible {\it iff} $r_{0a} =0$ \cite{LPS}. As
a consequence of \equ{000}, eq. \equ{rrr} reduces to
\be
0 = \b_g\ \Lp {1\over 128\, g^3} +
r_g\Rp +\sum_{\l_{ijk}}\b_{ijk}\,r_{ijk}\ .
\label{rrrred}
\ee
That the Yukawa couplings $\l_{ijk}$ are proportional to $g$ in
the one-loop approximation as a consequence of (iii) is clear
 from \equ{gamay}. At higher orders, $\l_{ijk}=\l_{ijk}(g)$ are
formal power series in $g$, and one needs to impose for
consistency that these functions satisfy the reduction
equations \equ{redred}.

A power series solution to the reduction equations exists at all
orders if there is a lowest-order solution which is unique.
We now look at this point in more details, following
\cite{LPS,Zim}.
At one-loop, the reduction equations \equ{redred} reduce to
\be
\b^{(1)}_{ijk}(\l_{ijk},g)=0\ ,\qquad\forall \l_{ijk}\ ,
\label{b1}
\ee
which is hypothesis (iv).
Separating the complex $\l_{ijk}$'s
into their real and imaginary parts,
we assume all Yukawa couplings to
be real and denote them by $\l_i$.
The reduction equations \equ{redred}
now read
\be
\b_g\ {d\l_i\over dg}~=~\b_i\ ,
\label{reds}
\ee
where the $\b$-functions have the forms \cite{Zim}
\be
\b_i
= \sum_{n=1}^\infty \sum_{a=0}^n \sum_k
C_{i}^{(n)\,k_1... k_{2a+1}}\,
g^{2n-2a}\,\l_{k_1}...\l_{k_{2a+1}}
= C_{i}^{(1)\,k}\,g^2\,\l_k + C_{i}^{(1)\,klm}\,
\l_k\l_l\l_m +\OO(\hbar^2)
\label{11}
\ee
with $n$ denoting the loop order, and
\be
\b_g
= g^3\,\sum_{n=2}^\infty \sum_{a=0}^{n-1} \sum_k
B^{(n)\,k_1... k_{2a}}\,
g^{2n-2-2a}\,\l_{k_1}...\l_{k_{2a}}
= \OO(\hbar^2)\ .
\label{22}
\ee
Having assumed hypothesis (ii), \ie $\b_g^{(1)} =0$, we look for a
solution of \equ{reds} of the form
\be
\l_i(g)=\sum_{n=0}^\infty \rho^{(n)}_{i}\,g^{2n+1}\ .
\label{33}
\ee
Inserting \equ{11}, \equ{22} and \equ{33} into \equ{reds}, one finds
 that $\rho^{(0)}_{i}$ must be a solution of the equations
\be
F_i[\rho^{(0)}]\equiv C_{i}^{(1)\,k}\,\rho_k^{(0)}
+C_{i}^{(1)\,klm}\,\rho_k^{(0)}\,\rho_l^{(0)}\,\rho_m^{(0)}=0\ ,
\label{b5}
\ee
which are just eqs. \equ{b1}, \ie hypothesis (iv).
At higher orders, one
 gets the recurrence conditions
\be
M^k_i\,\,\rho_k^{(n)}=f_i\ ,\qquad n\geq 1\ ,
\label{b6}
\ee
of which the right side depends only on $\rho^{(p)},\ p<n$. The
matrix $M$ depends on $\rho^{(0)}$ exclusively,
\be
M^k_i=\pad{F_i[\rho^{(0)}]}{\rho_k^{(0)}}\ .
\ee
If this matrix is non-singular, \ie {\it iff}
the solution $\rho^{(0)}$ of
\equ{b5} is unique,
then \equ{b6} determines
the higher order coefficients of \equ{33}
in terms of $\rho^{(0)}$.

Having done this, we replace the reduction equations
[in their form \equ{redred}] into \equ{rrrred} and get
\be
0 = \b_g\ \Lc {1\over 128\, g^3} + r_g + \sum_{\l_{ijk}}
{\pa\l_{ijk}\over \pa g}\,r_{ijk}\Rc\ ,
\label{rrrredrepl}
\ee
which is of the form $0=\b_g\,[1/128\,g^3+\OO(\hbar)]$, \ie
the bracket in \equ{rrrredrepl} is perturbatively invertible. It follows
that $\b_g=0$ at all orders, for the unique remaining
(independent) coupling of the theory, \eg the gauge coupling $g$.

%%%%%%%%%%%%%%%%%%%%%%%%%
%%%%%%%%%%%%%%%%%%%%%%%%%
%%%%%%%%%%%%%%%%%%%%%%%%%
\vspace{-3ex}
\section{A Realistic Finite SGUT Based on $SU(5)$}
\label{sec3}
\vspace{-2ex}
{}From the classification of
theories with $\beta_{g}^{(1)}=0$
\cite{HPS}, one can see that,
using $SU(5)$ as gauge group, there
exist only two candidate models which
can accommodate three fermion
generations. These models contain the
chiral supermutiplets
${\bf 5},~\overline{\bf 5},
{}~{\bf 10},~\overline{\bf 10},~{\bf 24}$
with the multiplicities $(6,9,4,1,0)$ and
 $(4,7,3,0,1)$, respectively.

Only the second model contains three families
and can describe in a self-consistent way
(\ie without reference to a larger model)
the spontaneous symmetry breaking (SSB) of $SU(5)$ down
to $SU(3)\times SU(2) \times U(1)$.
 We therefore concentrate here on that model, for which the
most general $SU(5)$ invariant, $N=1$ supersymmetric, cubic
superpotential is
\beq
W &=&{1\over 2} g_{ija}\ {\bf 10}_i {\bf 10} _j H_a
+ {\bar g}_{ija}\ {\bf 10}_i {\bar {\bf 5}}_j {\bar H}_a
+{1\over 2}{g^\prime}_{ijk}\ {\bf 10}_i
{\bar {\bf 5}}_j {\bar {\bf 5}}_k
+{1\over 2}q_{ia b}\ {\bf 10}_i {\bar H}_a {\bar H}_b\nn\\
&&+f_{a b}\ {\bar H}_a {\bf 24} H_b
+h_{i a}\ {\bar {\bf 5}}_i {\bf 24} H_a
+p\ {\bf 24}^3,
\label{superpot2}
\eeq
where $i,j,k=1,2,3$ and $a,b =1,2,3,4$; we have suppressed
the $SU(5)$ indices.
The ${\bf 10}_i$'s and ${\bar {\bf 5}}_i$'s
are the usual three generations,
and the ${\bf 24}$ contains the scalar superfield.
The four (${\bf 5}+{\bar {\bf 5}}$) Higgses are
denoted by $H_a$, ${\bar H}_a$.

Given the superpotential, the
$\gamma^{(1)}$'s can be easily computed
[$\beta_{g}^{(1)}$ vanishes of course].
Eq. (\ref{2nd}) imposes the following
relations among the Yukawa and
gauge couplings
\begin{equation}\begin{array}{rrcl}
\bar H:& 4\, {\bar g}_{ij a}\,
{\bar g}^{ijb} + {24\over 5}
\,f_{ac}\,f^{bc}+4\,q_{iac}\,q^{ibc}
\={24\over 5}\, g^2
 \,\delta _a ^b \ ,\\[1mm]
H:& 3\,g_{ija}\,g^{ijb}
+ {24\over 5}\, f_{ca}\,f^{cb}
+ {24\over 5}
\,h_{ia}\,h^{ib} \={24\over 5}
\, g^2 \,\delta _a ^b \ ,\\[1mm]
\bar {\bf 5}:& 4 \,{\bar g}_{ki a}
\,{\bar g}^{kja}+{24\over 5}
\,h_{ia}\,h^{ja}+4\,
{g^\prime}_{ikl}\,{g^\prime}^{jkl}
\={24\over 5}\, g^2 \,
\delta _i^j\ ,\\[1mm]
{\bf 10}:& 2 \,{\bar g}_{ik a}
\,{\bar g}^{jka}
+3\,g_{ika}\, g^{jka}+q_{iab}
\,q^{jab}+{g^\prime}_{kli}
\,{g^\prime}^{klj} \={36\over 5}
\, g^2 \,\delta _i ^j\ ,\\[1mm]
{\bf 24}:& f_{ab}\,f^{ab}+{21\over 5}
\, pp^*+ h_{ia}\,h^{ia}
\= 10\,g^2\ .
\label{yg}
\end{array}\end{equation}
To realize finiteness at all orders, we must find a
unique solution
of $\b^{(1)}_{ijk} =0$ [eq. \equ{411}], that is
consistent with the vanishing
of the $\gamma^{(1)}$'s [eqs. \equ{yg}].
Such a search contrasts with most of the previous studies of
the present model \cite{model1,model}, where
no attempt has been made to find isolated and non-degenerate
solutions. These studies have rather pursued an opposite goal.
They have used the freedom offered by the fact that the solutions
are not isolated in order to make specific {\it Ans{\" a}tze}
that could lead to phenomenologically acceptable
predictions.

Following \cite{kmz}, we concentrate on finding a
model (1) that is phenomenologically
interesting, \eg the $SU(5)$ SGUT
based on the (unconstrained) super\-poten\-tial
\equ{superpot2}, and (2) which yields a
solution to \equ{yg} that is unique as a solution
of \equ{411} in order to realize finiteness.
As a first approximation to the Yukawa
matrices, a diagonal solution (that is, one without
intergenerational mixing) may be considered.
It turns out that this can be achieved by imposing
the $Z_{7}\times Z_{3}$ discrete symmetry presented in
Table 1, plus a multiplicative $Q$-parity on the superpotential
$W$. Under the latter $Q$-parity, the
${\bf 10}_i$'s and $\overline{\bf 5}_i$'s describing
 the fermion multiplets are odd, while all other superfields
are even. These symmetries allow only $g_{iii},~\bar{g}_{iii},
{}~f_{44}$ and $p$ to be non-vanishing.
\begin{table*}[t]
\caption{The charges of the $Z_7\times Z_3$ symmetry}
\label{tbl}
$$
\begin{tabular}{|c|c|c|c|c|c|c|c|c|c|c|}
\hline
& ${\bf 10}_1$ & ${\bf 10}_2$ &
${\bf 10}_3$ &
 $\bar {\bf 5}_1$ & $\bar {\bf 5}_2
$ & $\bar {\bf 5}_3$& $H_1$ &
  $H_2$ & $H_3$ & $H_4$ \\ \hline
$Z_7$ &1 & 2  & 4 &  4  &
1  &  2  & 5&  3 & 6 & 0\\ \hline
$Z_3$ &1 &2   &0  & 0 &0
&0 &1 & 2 & 0 & 0\\ \hline
\end{tabular}
$$
\end{table*}
Furthermore, looking at this problem from the point of view of
the first formulation of the finiteness criterion (see Section
\ref{sec1}), we have found that there indeed exists a unique
power series solution of the reduction equations \equ{redred}
that satisfies the finiteness
conditions \equ{410}, \equ{411}. Defining
\be
\a_{ija}\equiv{|g_{ija}|^2\over 4\pi}\ ,\quad
\bar\a_{ija}\equiv{|\bar{g}_{ija}|^2\over 4\pi}\ ,\quad
\a_{f_{ab}}\equiv{|f_{ab}\,|^2\over 4\pi}\ ,\quad
\a_p\equiv{|p\,|^2\over 4\pi}\ ,\quad
{\rm and}\quad \a_{\rm GUT}
\equiv {|g\,|^2\over 4\pi}\ ,
\ee
this unique solution is given by \cite{kmz}
\be\ba{rclcrcl}
\alpha_{iii}\=\frac{8}{5}\,\alpha_{\rm GUT}
+\OO(\alpha_{\rm GUT}^2)\ ,\qq
\bar{\alpha}_{iii}\=\frac{6}{5}\,\alpha_{\rm GUT}
+\OO(\alpha_{\rm GUT}^2)\ ,\nn\\[2mm]
\alpha_{f_{44}}\=\alpha_{\rm GUT}
+\OO(\alpha_{\rm GUT}^2)~,\qq
\alpha_p\=\frac{15}{7}\,\alpha_{\rm GUT}
+\OO(\alpha_{\rm GUT}^2)~.
\label{solfut}
\ea
\ee
The $\OO(\alpha_{\rm GUT}^2)$-terms are power series in
$\alpha_{\rm GUT}$ that can be
uniquely computed to any finite order if the $\beta$-functions
of the unreduced model are known to the corresponding order.
The reduced model in which gauge and Yukawa couplings
are unified
has $\beta$-functions that identically vanish to that order.

In the above model, we have found
a diagonal solution for the Yukawa
couplings, with each family coupled to a different Higgs.
However, we may use the fact that mass terms
do not influence the $\beta$-functions in a certain
class of renormalization schemes, and introduce
appropriate mass terms that allow to rotate in the Higgs
sector so that only one pair of Higgs doublets, coupled to
the third family, remains light and acquires a
non-vanishing VEV (this is similar to \cite{model}).
Note that the effective coupling of the Higgs doublet
to the first family is very small, hence avoiding a
potential problem with proton lifetime \cite{proton}.
Thus, effectively, we have at low energies
the minimal supersymmetric standard model (MSSM) with
only one pair of Higgs doublets.

Supersymmetry breaking can be achieved through soft
breaking terms, which do not influence the
$\beta$-functions beyond $M_{\rm GUT}$.
It is worth mentioning that
renormalization group invariant relations in the
soft supersymmetry breaking sector
(which hold up to two loops) have been revived
\cite{jones2}. When these
conditions are applied to (a generalization of) the
present $SU(5)$ model, the finite case emerges
as the only possibility.

Since the $SU(5)$ symmetry is spontaneously broken
below $M_{\rm GUT}$, the finiteness conditions obviously
do not restrict the renormalization property at low energies.
All one gets at such regimes is a boundary condition on the
gauge and Yukawa couplings, which
have to be so chosen that they
satisfy (\ref{solfut}) at $M_{\rm GUT}$.
So we examine the evolution of the
gauge and Yukawa couplings according
to their renormalization group
equations at two-loops, taking into
account all the boundary conditions at
$M_{\rm GUT}$ \cite{kmz}.
A recent analysis \cite{top} based upon
updated experimental data
on Standard Model parameters
yields as a prediction for the top quark
mass in the present model
\be
m_t\ =\ (185\pm 5)\ {\rm GeV}\ .
\ee

%%%%%%%%%%%%%%%%%%%%%%%%
%%%%%%%%%%%%%%%%%%%%%%%%
%%%%%%%%%%%%%%%%%%%%%%%%
\vspace{-3ex}
\section{Conclusions}
\label{sec4}
\vspace{-2ex}
Electric-magnetic duality can be implemented in
supersymmetric Yang-Mills theories  if the
relations exchanging the weak and strong coupling
regimes can be made to hold at any renormalization
scale, \ie if the couplings do not run. This illustrates
the connection between electric-magnetic
duality and exact vanishing
of the $\b$-functions, \ie all-order perturbative finiteness.
We did not attempt here at giving a formal description
of that connection, which we have used mainly as a motivation
for discussing all-order finiteness.
Hoping to contribute to the exciting discussion
on duality symmetries, we have concentrated on the
case of all-order finite $N=1$ supersymmetric
gauge theories.

The criterion we have presented at the end of
Section \ref{sec2} for all-order vanishing $\b$-functions
is attractive due to the fact that it does involve only
one-loop hypotheses. The condition that the solutions
be isolated and non-degenerate [hypothesis (iv)] is
generally not met by an unconstrained model. This
does however not mean that SYM theories cannot
be made finite; finiteness can be achieved by enforcing
the unicity of such solutions through additional symmetry
requirements on the superpotential.

In general, the procedure of constructing an all-order finite
SYM theory involves two or more steps. One first reduces
 the number of independent Yukawa couplings by means
of global symmetries. Then one checks if the solution of
$\g^{i\ (1)}_{\ j} =0$ considered as a solution of
$\b^{(1)}_{ijk} = 0$ is isolated and non-degenerate. If not, the process has
 to be restarted, imposing an enlarged  global symmetry
to the superpotential. The process stops successfuly if
unicity of the solution of $\b^{(1)}_{ijk} = 0$ is attained.

One course, a more practical point of view may be adopted.
Starting from an unconstrained SYM model, one imposes the
global symmetries that are motivated by phenomenology
 (as, \eg family symmetry), and then checks if finiteness is
realized. One may hope that the global symmetries which
are necessary for finiteness turn out to be physically relevant
and to carry predictive power.

We have presented an application of the finiteness criterion
to a $SU(5)$ SGUT. The latter is shown to be all-order finite
provided one imposes discrete symmetries of the type
$Z_{7}\times Z_{3}$, plus a multiplicative $Q$-parity. The same
model yields a prediction for the top
quark mass, $m_t = (185\pm 5)$ GeV, which is in
agreement with the present experimental data.

%%%%%%%%%%%%%%%%%%%%%%%%
%%%%%%%%%%%%%%%%%%%%%%%%
%%%%%%%%%%%%%%%%%%%%%%%%
\vspace*{2ex}
{\bf Acknowledgements:}
The authors would like to thank those they have collaborated
with on the matters discussed in this paper: O. Piguet and K. Sibold (C.L.)
as well as D. Kapetanakis, J. Kubo, M. Mondragon and
M. Olechowski (G.Z.). We also express our gratitude to
J.-P. Derendinger and O. Piguet for helpful discussions
and useful comments on the manuscript.

%%%%%%%%%%%%%%%%%%%%%%%%
%%%%%%%%%%%%%%%%%%%%%%%%
%%%%%%%%%%%%%%%%%%%%%%%%

%%%%%%%%%%%%%%
\end{document}